\documentclass[runningheads]{llncs}
\usepackage[utf8]{inputenc}
\usepackage{times}
\usepackage[dvipsnames]{xcolor}
\usepackage{hyperref}
\usepackage{amssymb}
\usepackage{amsmath}
\usepackage{stmaryrd}

\newcommand{\draftonly}[1]{}

\usepackage{breakcites}

\usepackage{amsthm}

\usepackage{laws}

\usepackage{imakeidx}
\makeindex

\newcommand{\Hide}[2]{\exists_{#1}#2}

\makeatletter
\newcounter{colwidth}
\newenvironment{eqncolumns}{\@ifnextchar[{\@eqncolumns}{\@@eqncolumns}}{\end{eqnarray}\end{minipage}\vspace{1ex}}
\def\@eqncolumns[#1]{\setcounter{colwidth}{100-#1}\mbox{}\vspace{-4ex}\\\begin{minipage}[t]{0.#1\textwidth}\begin{eqnarray}}
\def\@@eqncolumns{\setcounter{colwidth}{5}\mbox{}\vspace{-4ex}\\\begin{minipage}[t]{0.5\textwidth}\begin{eqnarray}}
\newcommand{\secondcolumn}{\end{eqnarray}\end{minipage}\begin{minipage}[t]{0.\thecolwidth\textwidth}\begin{eqnarray}}
\makeatother

\newcommand{\existsallstates}{\reflectbox{$\mathsf{E}$}}
\newcommand{\existsfirst}{\reflectbox{$\mathsf{F}$}}
\newcommand{\existslast}{\reflectbox{$\mathsf{L}$}}
\newcommand{\HideSymS}{\mbox{\sc s}}
\newcommand{\HideS}[2]{\existsallstates^{\HideSymS}_{#1} #2}

\newcommand{\HideSymC}{\mbox{\sc c}}
\newcommand{\HideC}[2]{\existsallstates^{\HideSymC}_{#1} #2}
\newcommand{\HideFirstC}[2]{\existsfirst^{\HideSymC}_{#1} #2}
\newcommand{\HideLastC}[2]{\existslast^{\HideSymC}_{#1} #2}

\newcommand{\kw}[1]{\mathsf{#1}}
\newcommand{\invariant}{\mathop{\kw{inv}}}
\newcommand{\inv}[1]{\invariant #1}
\newcommand{\promise}[1]{\Om{(\cstep{#1})}} 
\newcommand{\CI}{\mathit{I}}
\newcommand{\prer}[1]{{}^\backprime#1}
\newcommand{\postr}[1]{#1{}^\prime}
\newcommand{\drefsto}[1]{\mathrel{\refsto_{#1}}}

\newcommand{\Eval}[2]{#1_{#2}}

\newcommand{\EqEvalC}[2]{\{ \sigma \spot \Eval{#1}{\sigma} = #2\}}

\newcommand{\Local}[2]{(\mathop{\kw{local}} #1 \spot #2)}
\newcommand{\variable}[2]{(\Var #1 \spot #2)}

\newcommand{\TLeventually}{\mathop{\lower0.0ex\hbox{$\Diamond$}}}
\newcommand{\TLfeventually}{\mathop{\lower0.0ex\hbox{$\blacklozenge$}}}
\newcommand{\TLalways}{\mathop{\lower0.0ex\hbox{$\Box$}}}
\newcommand{\TLfalways}{\mathop{\lower0.0ex\hbox{$\blacksquare$}}}

\newcommand{\Abort}{\lightning}

\newcommand{\Demand}{\mathop{\kw{guar}_{\boldsymbol{\epsilon}}}}
\newcommand{\Guarantee}{\mathop{\kw{guar}_{\boldsymbol{\pi}}}}

\newcommand{\Idle}{\kw{idle}}
\newcommand{\Rely}{\mathop{\kw{rely}}}
\newcommand{\Var}{\mathop{\kw{var}}}
\newcommand{\If}{\mathop{\kw{if}}}

\newcommand{\Then}{\mathbin{\kw{then}}}
\newcommand{\Else}{\mathbin{\kw{else}}}
\newcommand{\While}{\mathop{\kw{while}}}

\newcommand{\Do}{\mathbin{\kw{do}}}

\newcommand{\Expr}[2]{[\![#1]\!]_{#2}}

\newcommand{\True}{\mathsf{true}}
\newcommand{\False}{\mathsf{false}}
\newcommand{\Variant}{\mathop{\kw{variant}}}
\renewcommand{\implies}{\mathrel{\Rightarrow}}
\renewcommand{\iff}{\mathrel{\Leftrightarrow}}
\newcommand{\Comment}[1]{\mbox{{\color{purple}--- #1}}}
\newcommand{\relimplies}{\mathrel{\implies}}

\newcommand{\inter}{\mathbin{\cap}}
\newcommand{\union}{\mathbin{\cup}}
\newcommand{\spot}{\mathrel{.}}
\newcommand{\refsto}{\mathrel{\succeq}}

\newcommand{\nondet}{\mathbin{\vee}}
\newcommand{\Nondet}{\mathbin{\bigvee}}
\newcommand{\meet}{\mathbin{\wedge}}
\newcommand{\Meet}{\mathop{\bigwedge}}
\newcommand{\defs}{\mathrel{\widehat=}}

\newcommand{\cstepd}{\boldsymbol{\alpha}}
\newcommand{\cstep}[1]{\mathop{\alpha}#1}
\newcommand{\pstepd}{\pi}
\newcommand{\estepd}{\epsilon}
\newcommand{\cpstepd}{\boldsymbol{\pstepd}}
\newcommand{\cpstep}[1]{\mathop{\pstepd}#1}
\newcommand{\cestepd}{\boldsymbol{\estepd}}
\newcommand{\cestep}[1]{\mathop{\estepd}#1}
\newcommand{\cgd}[1]{\mathop{\tau}#1}
\newcommand{\Nil}{\boldsymbol{\tau}}
\newcommand{\Seq}{\mathbin{;}}
\newcommand{\Fin}[1]{#1^{\star}}
\newcommand{\Om}[1]{#1^{\omega}}
\newcommand{\Ata}{\mathsf{a}}

\newcommand{\together}{\mathbin{\Cap}}
\newcommand{\Pre}[1]{\{#1\}}

\newcommand{\rely}[1]{\Rely #1}
\newcommand{\guar}[1]{\Guarantee #1}

\newcommand{\demand}[1]{\Demand #1}
\newcommand{\Post}[1]{\Spec{}{}{#1}}
\newcommand{\atomicrel}[1]{\langle#1\rangle}
\newcommand{\opt}[1]{\mathop{\kw{opt}}#1}
\newcommand{\Term}{\kw{term}}
\newcommand{\dom}{\mathop{\kw{dom}}}

\makeatletter
\newcommand{\Frame}[2]{\ifx\@empty#1\else#1\!:\!\fi#2}
\def\Spec{\@ifnextchar*{\@Spec}{\@@Spec}}
\def\@Spec*#1#2#3{\ifx\@empty#1\else#1\colon\fi
   [{#2}\ifx\@empty#2\else,~\fi#3]}
\def\@@Spec#1#2#3{\ifx\@empty#1\else
   \begin{array}{@{}l@{}}#1\colon\end{array}\!\!\fi%
   \left[{\begin{array}{@{}l@{}}#2\end{array}}\ifx\@empty#2\else~,~~\fi
   \begin{array}{@{}l@{}}#3\end{array}\right]}
\makeatother
\newcommand{\id}[1]{{\textstyle\mathsf{id}}_{#1}}
\newcommand{\universalrel}{\mathsf{univ}}
\newcommand{\fun}{\mathbin{\rightarrow}}

\makeatletter
\def\Set{\@ifnextchar*{\@Set}{\@@Set}}
\def\@Set*#1{{\color{purple}\left\lfloor\begin{array}{l}#1\end{array}\right\rfloor}}
\def\@@Set#1{{\color{purple}\lfloor#1\rfloor}}
\def\Rel{\@ifnextchar*{\@Rel}{\@@Rel}}
\def\@Rel*#1{{\color{purple}\left\ulcorner\begin{array}{l}#1\end{array}\urcorner}}
\def\@@Rel#1{{\color{purple}\ulcorner#1\urcorner}} 
\def\RelA{\@ifnextchar*{\@RelA}{\@@RelA}}
\def\@RelA*#1{{\color{purple}\left\lceil\BB\left\lceil\begin{array}{l}#1\end{array}\right\rceil\BB\right\rceil}}
\def\@@RelA#1{{\color{purple}\lceil\BB\lceil#1\rceil\BB\rceil}}
\makeatother
\newcommand{\SPre}[1]{\Pre{{\color{purple}#1}}}

\newcommand{\RSpec}[3]{\Spec{#1}{}{\Rel{#3}}}

\newcommand{\Rrely}[1]{\rely{{\Rel{#1}}}}
\newcommand{\Rguar}[1]{\guar{{\Rel{#1}}}}
\newcommand{\Ratomicrel}[1]{\atomicrel{\Rel{#1}}}

\newcommand{\cat}{\mathbin{\raise 0.8ex\hbox{$\frown$}}}

\newcommand{\bool}{\mathbb{B}}

\newcommand{\lef}[1]{\mathrel{\leq_{#1}}}
\newcommand{\ltf}[1]{\mathrel{<_{#1}}}
\newcommand{\eqf}[1]{\mathrel{\equiv_{#1}}}

\makeatletter
\newcommand{\ChainRel}[1]{\crcr \noalign{\penalty\interdisplaylinepenalty}
  \hspace*{-1em}#1~ &
  \@ifnextchar*{\@ChainRelCommment}{}}
\newcommand{\Why}[1]{\mbox{{\color{blue}\hspace*{0em}#1}}}
\def\@ChainRelCommment*[#1]{\Why{#1}
  \crcr & 
  }
\newcommand{\StartRef}[1]{\hspace*{-1.5em} \ref{#1}) \refsto
  \@ifnextchar[{\@StartRefCommment}{}}
\def\@StartRefCommment[#1]{\mbox{#1}
  \crcr \noalign{\penalty\interdisplaylinepenalty}}
\makeatother

\newcommand{\Equiv}{\ChainRel{\equiv}}

\newcommand{\Refsto}{\ChainRel{\refsto}}

\newcommand{\Equals}{\ChainRel{=}}

\makeatletter
\def\@setmcodes#1#2#3{{\count0=#1 \count1=#3
  \loop \global\mathcode\count0=\count1 \ifnum \count0<#2
  \advance\count0 by1 \advance\count1 by1 \repeat}}

\DeclareSymbolFont{italic}{OT1}{\rmdefault}{m}{it}
\let\mathit\undefined
\DeclareSymbolFontAlphabet{\mathit}{italic}
\edef\@tempa{\hexnumber@\symitalic}
\@setmcodes{`A}{`Z}{"7\@tempa41}
\@setmcodes{`a}{`z}{"7\@tempa61}
\makeatother

\usepackage[colorinlistoftodos,textwidth=2cm]{todonotes}

\definecolor{CJ}{rgb}{1,1,0.9}
\definecolor{IH}{rgb}{1,0.9,1}
\definecolor{LM}{rgb}{0.9,1,1}

\newcounter{hours}
\newcounter{minutes}
\newcommand{\printtime}{%
  \ifthenelse{\value{hours}<10}{0}{}\thehours:%
  \ifthenelse{\value{minutes}<10}{0}{}\theminutes}

\newbox{\MyDate}
\savebox{\MyDate}{\draftonly{ (\today\ \printtime)}}

\title{Data reification in a\\ concurrent rely-guarantee algebra}
\titlerunning{Concurrent data reification\usebox{\MyDate}}

\author{
Larissa A. Meinicke\inst{1}\orcidID{0000-0002-5272-820X}
\and
Ian J. Hayes\inst{1}\orcidID{0000-0003-3649-392X}
\and
Cliff B. Jones\inst{2}\orcidID{0000-0002-0038-6623}
}

\institute{
School of Electrical Engineering and Computer Science, \\ 
The University of Queensland, Brisbane, Queensland 4072, Australia
\and
School of Computing Science, \\ 
Newcastle University, Newcastle upon Tyne, UK
  \draftonly{\\\vspace*{2ex} \today~\printtime}
}

\begin{document}

\maketitle

\begin{abstract}
Specifications of significant systems can be made short and perspicuous by using abstract data types; 
data reification can provide a clear, stepwise, development history of programs that use  
more efficient concrete representations. 
Data reification (or ``refinement'') techniques for sequential programs are well established.
This paper applies these ideas to concurrency,
in particular, an algebraic theory supporting rely-guarantee reasoning about concurrency.
A concurrent version of the Galler-Fischer equivalence relation data structure is used as an example.
\end{abstract}

\section{Introduction}\labelsect{introduction}

Data refinement techniques for sequential programs are well established \cite{Milner71a,Hoare72-data,Jones80a,Jones90a}.
(This paper mostly uses the verb ``reify'' and noun ``reification'' in the sense of making concrete.)
The specification of an abstract data structure (or type) includes both initialisation of the data structure
and a set of operations on the data structure.
The representation of the abstract data structure may use specification data types 
(e.g.~maps) that are not directly available in the target programming language
and the specification of the operations may be in terms of preconditions and postconditions, rather than code.
The data structure is encapsulated in the sense that it can only be accessed or modified using the operations defined on the data structure.
Because the data structure cannot be directly accessed by the program using it,
the representation used in an implementation of the data structure may differ from that in the specification
in order to provide a concrete data structure that is defined in terms of programming language data types
(e.g.~a hash table)
that provide efficient implementations of the abstract operations.

Let $v$ represent an abstract data structure with initialisation satisfying $iv$ and 
let $cv$ be a program using the abstract operations on $v$,
and let $w$ represent the corresponding implementation data structure with initialisation satisfying $iw$ 
and let $cw$ be the same as $cv$ but with any operation on $v$ within $cv$ replaced by the corresponding operation on $w$.
Both $v$ and $w$ can be types that are restricted by data type invariants.
Both $cv$ and $cw$ may use other variables in common that are not subject to a change of representation.
The overall correctness criteria for data reification is that any program running using the implementation of the data type 
is a refinement of the corresponding program using the abstract specification of the data type,
where $\Spec{v}{}{\postr{iv}}$ is a specification command that establishes the postcondition $iv$,
changing only variables in $v$.
\begin{align}
  \variable{v}{\Spec{v}{}{\postr{iv}} \Seq cv} & \refsto \variable{w}{\Spec{w}{}{\postr{iw}} \Seq cw}~. \labelprop{dr-correctness}
\end{align}

Data reification is a technique that allows this refinement to be justified by 
showing each abstract operation is refined by the corresponding concrete operation.
To do this a coupling invariant, $\CI$, is introduced relating the concrete and abstract representations.
Commonly, the coupling invariant is in the form of a data type invariant on the concrete state 
plus a ``retrieve'' function that extracts the abstraction representation from a concrete representation that satisfies the data type invariant.

For shared-memory concurrent programs, the main change is that $cv$ and hence $cw$ can include concurrency. 
The approach we take to handling reasoning about concurrent programs is 
based on the use of rely and guarantee conditions \cite{Jones81d,Jones83a,Jones83b,HJM-23}.
A rely condition of a thread $T$ is a binary relation $r$ between program states;
it represents an assumption that every transition from a state $\sigma_i$ to a state $\sigma_{i+1}$ made by the environment  of $T$
(i.e.\ all the threads running in parallel with $T$)
satisfies $r$, (i.e.\ $(\sigma_i,\sigma_{i+1}) \in r$).
Complementing this $T$ has a guarantee condition $g$, also a binary relation between states,
and $T$ ensures all program transitions it makes satisfy $g$, (i.e.\ $(\sigma_i,\sigma_{i+1}) \in g$).
The guarantee of $T$ must imply the rely conditions of all threads running in parallel with $T$.
Our semantics uses Aczel traces \cite{Aczel83,DaSMfaWSLwC}
that explicitly represent environment ($\estepd$) transitions as well as program ($\pstepd$) transitions (see \reffig{rely-guar}).
\begin{figure}[t]
\begin{center}
\input{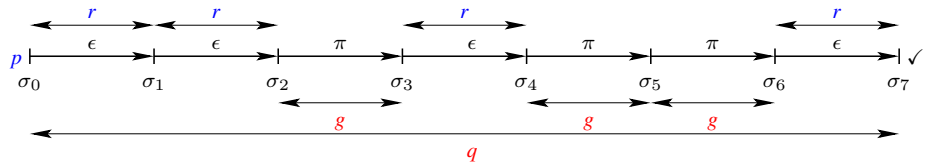}
\end{center}
\vspace{-4ex}
\caption{An Aczel trace satisfying a specification with precondition $p$, postcondition $q$, rely condition $r$ and guarantee condition $g$.
If the initial state, $\sigma_0$, is in $p$ and all environment transitions ($\estepd$) satisfy $r$,
then all program transitions ($\pstepd$) must satisfy $g$ and 
the initial state $\sigma_0$ must be related to the final state $\sigma_7$ by the postcondition $q$, also a relation between states.
}\labelfig{rely-guar}
\end{figure}

It was observed in~\cite{Jones06a} that data abstraction and reification can play an important role in formally developing concurrent programs.
The contribution of this paper is to extend data reification to concurrent programs based on the rely-guarantee style.
Our approach makes use of: 
\begin{itemize}
\item
program invariants that hold in every state and 
can be used to impose a coupling invariant between the abstract and concrete states 
\cite{2023MeinickeHayesDistributive-TR}
-- see \refsect{invariants},
and
\item
a data reification relation between commands defined in terms of a coupling invariant
-- see \refsect{dataref}.
\end{itemize}
\refsect{reify-var} uses these theories to show that the correctness condition \refprop{dr-correctness} is valid
and
\refsect{example} applies our theory to an interesting example: 
the development of the non-trivial representation of equivalence relations due to Galler and Fischer.
Our theory is supported by an extensive library of Isabelle/HOL theories \cite{IsabelleHOL}.
The next section introduces our wide-spectrum language.

\section{Wide-spectrum language}\labelsect{language}

\subsection{Language primitives}

Our language consists of a complete lattice of commands
with partial order, $c \refsto d$, meaning command $c$ is refined (or implemented) by command $d$,
so that non-deterministic choice ($c \nondet d$) is the lattice join 
and strong conjunction ($c \meet d$) is the lattice meet.
The lattice is complete so that for a set of commands $C$, 
non-deterministic choice $\Nondet C$ and strong conjunction $\Meet C$ are defined 
as the least upper bound and greatest lower bound, respectively.
The language includes binary operators for 
sequential composition ($c \Seq d$),
parallel composition ($c \parallel d$),
and 
weak conjunction ($c \together d$),
which behaves like strong conjunction ($\meet$) unless either operand aborts, in which case the weak conjunction aborts.

\paragraph{Naming and syntactic precedence conventions.}

We use 
$\sigma$ for program states (i.e.\ mappings from variable names to values),
$c$ and $d$ for commands;
$p$ and $I$ for sets of program states (with $I$ reserved for invariants);
$g$, $q$ and $r$ for binary relations between program states;
$e$ for expressions,
and 
$k$ for values.
Subscripted versions of the above names follow the same convention.
Unary operations and function application have higher precedence than binary operations.
Non-deterministic choice ($\nondet$) has the lowest precedence, 
and sequential composition ($\Seq$) has the highest precedence,
except framing ($:$) is higher.
We use parentheses to resolve all other syntactic ambiguities.

A program state $\sigma \in \Sigma$ gives the values of the program variables.
The language includes four primitive commands:
\begin{description}
\item[$\cgd{p}$]
is an instantaneous test that the current state $\sigma$ satisfies $p$:
if $\sigma \in p$ it is a no-op, otherwise it is infeasible;
\item[$\cpstep{r}$]
is an atomic program command that may perform a program transition from $\sigma_0$ to $\sigma$ if $(\sigma_0,\sigma) \in r$,
otherwise it is infeasible if $\sigma \not\in \dom r$;
\item[$\cestep{r}$]
is an atomic environment command that may perform an environment transition from $\sigma_0$ to $\sigma$ if $(\sigma_0,\sigma) \in r$,
otherwise it is infeasible if $\sigma \not\in \dom r$;
\item[$\Abort$]
is Dijkstra's abort command \cite{Dijkstra75,Dijkstra76} that irrecoverably fails immediately.
It is the greatest command.
\end{description}
The command $\Nil$ is the test that always succeeds \refdef{nil}.
The assertion command, $\Pre{p}$, aborts if the current state is not in $p$, where $\overline{p} = \Sigma - p$, otherwise it is a no-op \refdef{pre}.
The command $\cstep{r}$ allows either a program or an environment transition, provided it satisfies $r$ \refdef{cstep}.
The abbreviations $\cpstepd$ and $\cestepd$ allow any program \refdef{cpstepd} or environment \refdef{cestepd} transition, respectively,
and $\cstepd$ allows any transition, program or environment \refdef{cstepd},
where $\universalrel$ is the universal relation between program states.
Note the bold fonts for $\Nil$, $\cpstepd$, $\cestepd$ and $\cstepd$.
\begin{eqncolumns}
  \Nil & \defs & \cgd{\Sigma} \labeldef{nil} \\
  \Pre{p} & \defs & \Nil \nondet \cgd{\overline{p}} \Seq \Abort \labeldef{pre} \\
  \cstep{r} & \defs & \cpstep{r} \nondet \cestep{r} \labeldef{cstep}
\secondcolumn
  \cpstepd & \defs & \cpstep{\universalrel} \labeldef{cpstepd} \\  
  \cestepd & \defs & \cestep{\universalrel} \labeldef{cestepd} \\  
  \cstepd & \defs & \cstep{\universalrel} \labeldef{cstepd}  
\end{eqncolumns}
The basic commands satisfy the following refinement properties.
\begin{eqncolumns}
  \cgd{p_1} & \refsto & \cgd{p_2}  ~~~~~\mbox{if } p_1 \supseteq p_2 \labelprop{cgd-ref} \\
  \Pre{p_1} & \refsto & \Pre{p_2}  ~~~~\mbox{if } p_1 \subseteq p_2 \labelprop{pre-ref} 
\secondcolumn
  \cpstep{r_1} & \refsto & \cpstep{r_2}  ~~~~\mbox{if } r_1 \supseteq r_2 \labelprop{cpstep-ref} \\
  \cestep{r_1} & \refsto & \cestep{r_2}  ~~~~\mbox{if } r_1 \supseteq r_2 \labelprop{cestep-ref} 
\end{eqncolumns}
A command, $\Ata$, is considered \emph{atomic} if it is of the form $\Ata = \cpstep{g} \nondet \cestep{r}$ for some relations $g$ and $r$.
That is, $\Ata$ can only make a single transition, 
which may be a program ($\pstepd$) transition in $g$ or an environment ($\estepd$) in $r$.
A test can be pulled out of a branch of a weak conjunction provided the other branch does not immediately abort,
(i.e.\ $d$ starts with an atomic command or $d$ is a test).
\begin{align}
  \cgd{p} \Seq c \together d & = \cgd{p} \Seq (c \together d)  && \mbox{if $d$ is not immediately aborting} \labelprop{test-command-sync-command}
\end{align}

\subsection{Derived commands}

Finite iteration, $\Fin{c}$, and possibly infinite iteration, $\Om{c}$, of a command $c$
are defined as the least and greatest fixed points, respectively, of the function $(\lambda y \spot \Nil \nondet c \Seq y)$.
A program guarantee command, $\guar{r}$ for relation $r$, requires every program transition satisfies $r$
but places no constraints on environment transitions \refdef{guar}.
An environment guarantee command, $\demand{r}$, requires every environment transition satisfies $r$ \refdef{demand}.
A rely command, $\rely{r}$, assumes environment transitions satisfy $r$;
if one does not it aborts \refdef{rely},
in the same way that an assertion $\Pre{p}$ aborts if the initial state is not in $p$.
The command $\Term$ only performs a finite number of program transitions
but does not constrain its environment \refdef{term}.
The relation $\id{}$ stands for the identity relation and
$\id{v}$, is the identity function on a set of variables $v$,
and $\overline{v}$ is the set complement of $v$.
The command $\Frame{v}{c}$, restricts $c$ so that its program transitions may only modify variables in the set $v$ \refdef{frame}.
Within the identity function and frames a variable $u$ stands for the singleton set $\{u\}$ and a comma is used for union,
so that for variables $u$ and $w$, $\id{u,w} = \id{\{u\} \union \{w\}} = \id{\{u,w\}}$.
Similarly, $\id{\overline{u}} = \id{\overline{\{u\}}}$.
The command $\Idle$ can perform only a finite number of idling (no change) program transitions
but does not constrain its environment \refdef{idle}.
The command $\opt{r}$ either performs a single program transition satisfying $r$
or it can terminate immediately from states $\sigma$ such that $(\sigma,\sigma) \in r$ \refdef{opt}.
The atomic specification command, $\atomicrel{r}$, 
achieves $r$ via an (atomic) program transition 
and allows finite stuttering program transitions (that do not change the state) before and after 
the atomic transition \refdef{atomic-spec};
like $\opt{r}$, the atomic transition satisfying $r$ may be elided for states $\sigma$ in which $(\sigma,\sigma) \in r$. 
\begin{eqncolumns}[40]
  \Fin{c} & \defs & \mu y \spot \Nil \nondet c \Seq y \labeldef{finite-iter} \\
  \Om{c} & \defs & \nu y \spot \Nil \nondet c \Seq y \labeldef{iter} \\
  \guar{r} & \defs & \Om{(\cpstep{r} \nondet \cestepd)}  \labeldef{guar} \\
  \demand{r} & \defs & \Om{(\cpstepd \nondet \cestep{r})} \labeldef{demand} \\
  \rely{r} & \defs & \Om{(\cstepd \nondet \cestep{\overline{r}} \Seq \Abort)}  \labeldef{rely}
\secondcolumn
  \Term & \defs & \Fin{\cstepd} \Seq \Om{\cestepd}  \labeldef{term} \\
  \Frame{v}{c} & \defs & \guar{\id{\overline{v}}} \together c \labeldef{frame} \\
  \Idle & \defs & \Frame{\emptyset}{\Term}  \labeldef{idle} \\
  \opt{r} & \defs & \cpstep{r} \nondet \cgd{\{\sigma \spot (\sigma,\sigma) \in r\}}  \labeldef{opt} \\
  \atomicrel{r} & \defs & \Idle \Seq \opt{r} \Seq \Idle  \labeldef{atomic-spec} 
\end{eqncolumns}\\
The postcondition specification command, $\Post{q}$, for a relation between states $q$,
guarantees to terminate in a final state $\sigma$ that is related to the initial state $\sigma_0$ by $q$ \refdef{spec}.
The command $\Expr{e}{k}$ succeeds if the expression $e$ evaluates to the value $k$ but becomes infeasible otherwise.%
\footnote{See \cite{hayes2021deriving} for the complete details of expressions.}
An assignment command, $v := e$, 
is non-deterministic due to possible interference from concurrent threads modifying the values of the variables used within $e$.
The notation $\sigma[v \mapsto k]$ stands for $\sigma$ updated so that $v$ maps to $k$.
An assignment evaluates $e$ to some value $k$, which is then atomically assigned to $v$ \refdef{assignment};
the $\Idle$ at the end allows for hidden stuttering steps in the implementation.
A conditional either evaluates $b$ to $true$ and executes $c$, or to $false$ and executes $d$ \refdef{conditional};
the $\Idle$ command at the end allows for (hidden) program branching transitions.
A while loop executes $c$ while $b$ evaluates to $true$ and terminates when $b$ evaluates to $false$ \refdef{while}.
\begin{eqnarray}
  \Spec{}{}{q} &\defs & \Nondet_{\sigma_0} \cgd{\{\sigma_0\}} \Seq \Term \Seq \cgd{\{\sigma \spot (\sigma_0,\sigma) \in q\}}  \labeldef{spec} \\
  v := e & \defs & \Nondet_k \Expr{e}{k} \Seq \Frame{v}{\opt{\{(\sigma_0,\sigma) \spot \sigma = \sigma_0[v \mapsto k]\}}} \Seq \Idle \labeldef{assignment} \\
  \If b \Then c \Else d & \defs & (\Expr{b}{\True} \Seq c \nondet \Expr{b}{\False} \Seq d) \Seq \Idle \labeldef{conditional} \\
  \While b \Do c & \defs & \Om{(\Expr{b}{\True} \Seq c)} \Seq \Expr{b}{\False} \labeldef{while}
\end{eqnarray}

\subsection{Local variable blocks}\labelsect{local}

Local variable blocks are defined in terms of a more primitive operators 
akin to existential quantification of a variable over a command.
The algebra of these operators is based on Tarski's cylindric algebra%
\footnote{The name ``cylindric'' is appropriate for their model but not really appropriate for our model.}
\cite{HenkinMonkTarski71},
which introduces an existential operator $\Hide{v}{x}$ for each variable $v$ and variable and element $x$.
An element $x$ is \emph{independent} of $v$ if $x = \Hide{v}{x}$.
A standard model of their algebra is sets of states, 
and for a set of states $p$,
$\HideS{v}{p}$ is the set of states $\sigma$ such that there exists a state $\sigma_p \in p$,\
such that $\sigma$ and $\sigma_p$ are equal except possibly for the value of $v$.

Our earlier work extended cylindric algebra to apply to commands for sequential programs \cite{MPC19CylindricAlgebra}
and concurrent programs \cite{FormaliSE_localisation}, from which the properties used here are taken.
We make use of two existential operators on commands:%
\footnote{There is also first state existential operator, $\HideFirstC{v}{c}$, but we do not need that here.}
\begin{description}
\item[$\HideC{v}{c}$]
that has a trace $t$ if there exists a trace $t_c$ of $c$ such that the two traces are the same except for the value of $v$ in each state,
\item[$\HideLastC{v}{c}$]
that has a trace $t$ if there exists a trace $t_c$ of $c$ such that the two traces are the same except for the value of $v$ in the final state;
if $t$ is infinite (i.e.\ there is no final state) then $t$ must be a trace of $c$.
\end{description}
The following property, in which $\Ata$ is an atomic command and $c$ is a command,
allows an existential to be distributed over a sequential composition.
\begin{align}
  \HideC{v}{(c \Seq \Om{\Ata})} & = \HideC{v}{c}\Seq \Om{(\HideC{v}{\Ata})} &&
  \mbox{if } \HideLastC{v}{\Ata} = \HideC{v}{\Ata} \labelprop{hide-seq-iter-atomic}
\end{align}
A local block introduces a local variable $v$ \refdef{local}.
If $v$ exists as a variable in the outer context of the local block, its value is unmodified by program transitions of the local block 
(i.e.\ $\guar{\id{v}}$).
The variable $v$ is existentially quantified over the scope of the local block,
and within that block, $v$ cannot be modified by environment transitions (i.e.\ $\demand{\id{v}}$).
A variable block \refdef{variable} introduces a local scope for $v$ 
but includes $\Idle$ commands before and after to represent hidden actions to allocate and deallocate $v$.
\begin{align}
  \Local{v}{c} & \defs \guar{\id{v}} \together \HideC{v}{(\demand{\id{v}} \together c)} \labeldef{local} \\
  \variable{v}{c} & \defs \Idle \Seq \Local{v}{c} \Seq \Idle \labeldef{variable}
\end{align}
A command $c$ is \emph{independent} of a variable $v$ if $c = \HideC{v}{c}$.
The following properties follow from the theory in \cite{FormaliSE_localisation}.
\begin{align}
  \Frame{u}{c} & = \Local{v}{\Frame{v,u}{c}} && \mbox{if } u \neq v \mbox{ and } c = \HideC{v}{c} \labelprop{intro-local-framed} \\
  \variable{v}{\Local{w}{c}} & = \variable{w}{\Local{v}{c}} \labelprop{interchange-vars} \\
  \Local{v}{(\Frame{v,u}{c}) \Seq d} & = \Frame{u}{c} \Seq \Local{v}{d} && \mbox{if } u \neq v \mbox{ and } c = \HideC{v}{c} \labelprop{restrict-local} \\
  \Local{v}{c \together d} & = \Local{v}{c} \together d && \mbox{if } d = \HideC{v}{d} \labelprop{local-conj-indep} \\
  \Local{v}{c \together (\Frame{v,u}{d})} & = \Local{v}{c} \together \Frame{u}{d} && \mbox{if } u \neq v \mbox{ and } d = \HideC{v}{d} \labelprop{local-conj-indep-framed}
\end{align}

\section{Invariants}\labelsect{invariants}

Given a set of states $p$, we define $\prer{p}$ and $\postr{p}$
to be the relations between initial and final program states that
satisfy $p$ in their initial and final program states, respectively.
A pair $(\sigma_0, \sigma)$ is in relation $r_1 \relimplies r_2$, if it is not in $r_1$ or it is in $r_2$.
It is exclusively used here in the form $\prer{p} \relimplies \postr{p}$,
that is, the set of pairs $(\sigma_0,\sigma)$ such that either $\sigma_0 \notin p$ or $\sigma \in p$.
\begin{eqncolumns}
  \prer{p} & \defs \{ (\sigma_0,\sigma) \mid \sigma_0 \in p \} \labeldef{postr} \\
  \postr{p} & \defs \{ (\sigma_0, \sigma) \mid \sigma \in p \} \labeldef{prer} 
\secondcolumn
  r_1 \relimplies r_2 & \defs \overline{r_1} \union r_2 \labeldef{relimplies}
\end{eqncolumns}
The command $\inv{I}$ can take an arbitrary number of atomic transitions,
each of which establishes $I$ in its final state, assuming the initial state satisfies $I$.

\begin{definitionx}[inv]
For a set of states $I$,
\(
  \inv{I} \defs  \Pre{I} \Seq \promise{\postr{I}} 
\)
\end{definitionx}

An invariant can be introduced if it holds initially \refprop{inv-introduce}.
An invariant maintains $I$ on every transition \refprop{inv-maintains}.
\begin{align}
  \Pre{I} \Seq c & \refsto \inv{I} \together c \labelprop{inv-introduce} \\
  \inv{I} & = \Pre{I} \Seq \Om{(\cstep{(\prer{I} \relimplies \postr{I})})} \labelprop{inv-maintains} 
\end{align}

\begin{lemmax}[inv-distribute]
\begin{align}
  \inv{I} \together \Nondet C & = \Nondet_{c \in C} (\inv{I} \together c) && \mbox{if } C \neq \emptyset \labelprop{inv-distrib-Nondet} \\
  \inv{I} \together (c \nondet d) & = (\inv{I} \together c) \nondet (\inv{I} \together d)  \labelprop{inv-distrib-nondet} \\
  \inv{I} \together (c \Seq d) & = (\inv{I} \together c) \Seq (\inv{I} \together d)  \labelprop{inv-distrib-seq} \\
  \inv{I} \together (c \parallel d) & = (\inv{I} \together c) \parallel (\inv{I} \together d)  \labelprop{inv-distrib-par} \\
  \inv{I} \together (c \together d) & =  (\inv{I} \together c) \together (\inv{I} \together d) \labelprop{inv-distrib-conj} \\
  \inv{I} \together \Fin{c} & = \Fin{(\inv{I} \together c)} \labelprop{inv-distrib-finite-iter} \\
  \inv{I} \together \Om{c} & = \Om{(\inv{I} \together c)} \labelprop{inv-distrib-iter} \\
  \inv{I} \together \cgd{p} & = \Pre{I} \Seq \cgd{p}  \labelprop{inv-test} \\
  \inv{I} \together \Pre{p} & = \Pre{I \inter p}  \labelprop{inv-assert} \\
  \inv{I} \together \cpstep{r} & = \Pre{I} \Seq \cpstep{(r \inter \postr{I})}  \labelprop{inv-cpstep} \\
  \inv{I} \together \cestep{r} & = \Pre{I} \Seq \cestep{(r \inter \postr{I})}  \labelprop{inv-cestep} 
\end{align}
\end{lemmax}

\section{Data reification under coupling a coupling invariant}\labelsect{dataref}

We say that a command $d$ is a data reification of a command $c$ under coupling
invariant $\CI$, written $c \drefsto{\CI} d$, when $c$ is reified by
$d$ under the assumption that the coupling invariant $\CI$ is
initially true and $\CI$ is maintained by every atomic step.
\begin{definitionx}[data-refines]
For coupling invariant $\CI$ and commands $c$ and $d$,
\begin{align*}
 c \drefsto{\CI} d & \defs (\inv{\CI} \together c  \refsto  \inv{\CI} \together d) 
\end{align*}
\end{definitionx}
From the definition, data reification is a preorder,
that is, it is reflexive and transitive,
and $c \drefsto{\CI} d$ if $c \refsto d$.
Reifying a composite command can be reduced to reifying its components.
\begin{lemmax}[reify-operators]
\begin{align}
  \Nondet C & \drefsto{\CI} \Nondet D && \mbox{if } \forall d \in D \spot \exists c \in C \spot c \drefsto{\CI} d \labelprop{reify-Nondet} \\
  c_1 \nondet c_2 & \drefsto{\CI} d_1 \nondet d_2 && \mbox{if } c_1 \drefsto{\CI} d_1 \mbox{ and } c_2 \drefsto{\CI} d_2 \labelprop{reify-nondet} \\
  c_1 \Seq c_2 & \drefsto{\CI} d_1 \Seq d_2 && \mbox{if } c_1 \drefsto{\CI} d_1 \mbox{ and } c_2 \drefsto{\CI} d_2 \labelprop{reify-seq} \\
  c_1 \parallel c_2 & \drefsto{\CI} d_1 \parallel d_2 && \mbox{if } c_1 \drefsto{\CI} d_1 \mbox{ and } c_2 \drefsto{\CI} d_2 \labelprop{reify-par} \\
  c_1 \together c_2 & \drefsto{\CI} d_1 \together d_2 && \mbox{if } c_1 \drefsto{\CI} d_1 \mbox{ and } c_2 \drefsto{\CI} d_2 \labelprop{reify-conj} \\
  \Fin{c} & \drefsto{\CI} \Fin{d} && \mbox{if } c \drefsto{\CI} d \labelprop{reify-finite-iter} \\
  \Om{c} & \drefsto{\CI} \Om{d} && \mbox{if } c \drefsto{\CI} d \labelprop{reify-iter} 
\end{align}
\end{lemmax}

\begin{proof}
In each case the definition $\drefsto{\CI}$ is unfolded to make use of the invariant command,
the invariant is distributed over the operator using (\refprop*{inv-distrib-Nondet}--\refprop*{inv-distrib-iter}),
data reification of the component commands follows from the corresponding assumption,
the distribution of the invariant is reversed,
and
the result is folded using the definition of $\drefsto{\CI}$.
\end{proof}

\begin{lemma}[reify-commands]
\begin{align}
  \cgd{p_1} & \drefsto{\CI} \cgd{p_2} && \mbox{if } p_1 \supseteq \CI \inter p_2\labelprop{reify-test} \\
  \cpstep{r_1} & \drefsto{\CI} \cpstep{r_2} && \mbox{if } r_1 \supseteq \prer{\CI} \inter r_2 \inter \postr{\CI} \labelprop{reify-cpstep} \\
  \cestep{r_1} & \drefsto{\CI} \cestep{r_2} && \mbox{if } r_1 \supseteq \prer{\CI} \inter r_2 \inter \postr{\CI} \labelprop{reify-cestep} \\
  \Pre{p_1} & \drefsto{\CI} \Pre{p_2} && \mbox{if } \CI \inter p_1 \subseteq p_2 \labelprop{reify-assert} \\
  \guar{g_1} & \drefsto{\CI} \guar{g_2} && \mbox{if } g_1 \supseteq \prer{\CI} \inter g_2 \inter \postr{\CI} \labelprop{reify-guar} \\
  \rely{r_1} & \drefsto{\CI} \rely{r_2} && \mbox{if } \prer{\CI} \inter r_1 \inter \postr{\CI} \subseteq  r_2 \labelprop{reify-rely} \\
  \Frame{w}{c} & \drefsto{\CI} \guar{g} \together \Frame{v,w}{c} && \mbox{if } \id{v} \supseteq \prer{\CI} \inter g \inter \postr{\CI} \mbox{ and } v \neq w \labelprop{reify-frame} \\
  \opt{q_1} & \drefsto{\CI} \opt{q_2} && \mbox{if } q_1 \supseteq \prer{\CI} \inter q_2 \inter \postr{\CI} \labelprop{reify-opt} \\
  \Pre{p} \Seq \Post{q_1} & \drefsto{\CI} \Pre{\CI \inter p} \Seq \Post{q_2} && \mbox{if }  q_1 \supseteq \prer{\CI} \inter \prer{p} \inter q_2 \inter \postr{\CI} \labelprop{reify-spec}
\end{align}

\end{lemma}

\begin{proof}
For \refprop{reify-test}, expanding \Definition{data-refines}, we must show,
\(
  \inv{\CI} \together \cgd{p_1} \refsto \inv{\CI} \together \cgd{p_2} ,
\)
or combining invariant and test by \refprop{inv-test},
\(
  \Pre{\CI} \Seq \cgd{p_1} \refsto \Pre{\CI} \Seq \cgd{p_2} ,
\)
which follows by strengthening the test under the precondition, $\CI$, as $p_1 \supseteq \CI \inter p_2$.

We give the proof for \refprop{reify-cpstep}, the proof for \refprop{reify-cestep} is similar.
Expanding \Definition{data-refines}, we must show,
\(
  \inv{\CI} \together \cpstep{r_1} \refsto \inv{\CI} \together \cpstep{r_2},
\)
or combining invariant and program transition by \refprop{inv-cpstep},
\(
  \Pre{\CI} \Seq \cpstep{(r_1 \inter \postr{\CI})} \refsto \Pre{\CI} \Seq \cpstep{(r_2 \inter \postr{\CI})},
\)
which follows by strengthening the relation under the context of the precondition because $r_1 \supseteq \prer{\CI} \inter r_2 \inter \postr{\CI}$.

For \refprop{reify-assert}, by \refprop{inv-assert},
$\inv{\CI} \together \Pre{p_1} = \Pre{\CI \inter p_1} \refsto \Pre{\CI \inter p_2} = \inv{\CI} \together \Pre{p_2}$,
as the assumption implies $\CI \inter p_1 \subseteq \CI \inter p_2$.

For \refprop{reify-guar}, by \refprop{reify-nondet} and 
\refprop{reify-cpstep} using assumption $g_1 \supseteq \prer{\CI} \inter g_2 \inter \postr{\CI}$ for the program transition,
\(
  (\cpstep{g_1} \nondet \cestepd) \drefsto{\CI} (\cpstep{g_2} \nondet \cestepd) ,
\)
and hence by \refprop{reify-iter},
\(
  \Om{(\cpstep{g_1} \nondet \cestepd)} \drefsto{\CI} \Om{(\cpstep{g_2} \nondet \cestepd)} ,
\)
and the result follows by the definition of a guarantee \refdef{guar}.

For \refprop{reify-rely}, $\overline{r_1} \supseteq \prer{\CI} \inter \overline{r_2} \inter \postr{\CI}$ 
is equivalent to the assumption $\prer{\CI} \inter r_1 \inter \postr{\CI} \subseteq  r_2$:
\begin{align*}&
  \overline{r_1} \supseteq \prer{\CI} \inter \overline{r_2} \inter \postr{\CI}  
 \iff
  r_1 \subseteq \overline{\prer{\CI} \inter \overline{r_2} \inter \postr{\CI}}
 \iff
  r_1 \subseteq \prer{\overline{\CI}} \union r_2 \union \postr{\overline{\CI}}
 \iff
  \prer{\CI} \inter r_1 \inter \postr{\CI} \subseteq r_2 .
\end{align*}%
By \refprop{reify-nondet}, \refprop{reify-seq} and 
\refprop{reify-cestep} using $\overline{r_1} \supseteq \prer{\CI} \inter \overline{r_2} \inter \postr{\CI}$ for the environment transition,
\[
  (\cpstepd \nondet \cestepd \nondet \cestep{\overline{r_1}} \Seq \Abort) \drefsto{\CI} (\cpstepd \nondet \cestepd \nondet \cestep{\overline{r_2}} \Seq \Abort)
\]
and hence by \refprop{reify-iter},
\(
  \Om{(\cpstepd \nondet \cestepd \nondet \cestep{\overline{r_1}} \Seq \Abort)} \drefsto{\CI} \Om{(\cpstepd \nondet \cestepd \nondet \cestep{\overline{r_2}} \Seq \Abort)} ,
\)
and the result follows by the definition of a rely \refdef{rely}.

For \refprop{reify-frame}, the definition of framing \refdef{frame} is expanded to a guarantee,
which is then split,
$\Frame{w}{c} 
= \guar{\id{\overline{w}}} \together c
= \guar{\id{v}} \together \guar{\id{\overline{v,w}}} \together c
= \guar{\id{v}} \together \Frame{v,w}{c}
\drefsto{\CI} \guar{g} \together \Frame{v,w}{c}
$
by \refprop{reify-guar} using the assumption $\id{v} \supseteq \prer{\CI} \inter g \inter \postr{\CI}$.

For \refprop{reify-opt}, the proof follows from the definition of $\opt{q_1}$ \refdef{opt},
\refprop{reify-nondet}, \refprop{reify-cpstep} and \refprop{reify-test}.
\begin{align*}&
  \opt{q_1}
 =
  \cpstep{q_1} \nondet \cgd{\{\sigma \spot (\sigma,\sigma) \in q_1\}}
 \drefsto{\CI}
  \cpstep{q_2} \nondet \cgd{\{\sigma \spot (\sigma,\sigma) \in q_2\}}
 =
  \opt{q_2}
\end{align*}%
For the application of \refprop{reify-test} one needs to show 
$\{\sigma \spot (\sigma,\sigma) \in q_1\} \supseteq \CI \inter \{\sigma \spot (\sigma,\sigma) \in q_2\}$, 
given the assumption, $q_1 \supseteq \prer{\CI} \inter q_2 \inter \postr{\CI}$.
\begin{align*}
  \{\sigma \spot (\sigma,\sigma) \in q_1\}
&  \supseteq 
  \{\sigma \spot (\sigma,\sigma) \in \prer{\CI} \inter q_2 \inter \postr{\CI}\}
 =
  \{\sigma \spot (\sigma,\sigma) \in q_2 \land \sigma \in \CI\} \\
 & =
  \CI \inter \{\sigma \spot (\sigma,\sigma) \in q_2\}
\end{align*}
For \refprop{reify-spec}, the result follows from the definition of a specification command \refdef{spec},
refining using the assumption $ q_1 \supseteq \prer{\CI} \inter \prer{p} \inter q_2 \inter \postr{\CI}$,
reifying both the assertion \refprop{reify-assert} and the test \refprop{reify-test},
noting the initial state is assumed to be in $\CI \inter p$,
and folding using the definition of a specification command \refdef{spec}.
\begin{align*}
  \Pre{p} \Seq \Post{q_1} 
& =
  \Pre{p} \Seq \Nondet_{\sigma_0} \cgd{\{\sigma_0\}} \Seq \Term \Seq \cgd{\{ \sigma \spot (\sigma_0,\sigma) \in q_1\}} \\
& \refsto
  \Pre{p} \Seq \Nondet_{\sigma_0} \cgd{\{\sigma_0\}} \Seq \Term \Seq \cgd{\{ \sigma \spot (\sigma_0,\sigma) \in \prer{\CI} \inter \prer{p} \inter q_2 \inter \postr{\CI} \}} \\
&  \drefsto{\CI}
  \Pre{\CI \inter p} \Seq \Nondet_{\sigma_0} \cgd{\{\sigma_0\}} \Seq \Term \Seq \cgd{\{ \sigma \spot (\sigma_0,\sigma) \in \prer{\CI} \inter \prer{p} \inter q_2\}} \\
&  =
  \Pre{\CI \inter p} \Seq \Nondet_{\sigma_0} \cgd{\{\sigma_0\}} \Seq \Term \Seq \cgd{\{ \sigma \spot (\sigma_0,\sigma) \in q_2\}} \\
&  =
  \Pre{\CI \inter p} \Seq \Post{q_2}
 \qedhere
\end{align*}
\end{proof}

\begin{lemmax}[reify-rgspec]
If
$\CI \inter p_1 \subseteq p_2$ and
$\prer{\CI} \inter r_1 \inter \postr{\CI} \subseteq r_2$ and
$g_1 \supseteq \prer{\CI} \inter g_2 \inter \postr{\CI}$ and
$q_1 \supseteq \prer{\CI} \inter \prer{p_1} \inter q_2 \inter \postr{\CI}$
then
\[
  \rely{r_1} \together \guar{g_1} \together \Pre{p_1} \Seq \Post{q_1}
 \drefsto{\CI} 
  \rely{r_2} \together \guar{g_2} \together \Pre{p_2} \Seq \Post{q_2} .
\]
\end{lemmax}

\begin{proof}
By \refprop{reify-conj}, \refprop{reify-rely}, \refprop{reify-guar}, \refprop{reify-spec}.
\end{proof}

The notation 
$e_{\sigma}$ stands for the value of $e$ in state $\sigma$.
If an expression $e$ is single-reference under a rely condition $r$,
evaluating $e$ under interference satisfying $r$ corresponds to 
evaluating $e$ in one of the states during its execution,
and hence the following property holds.%
\footnote{See \cite{hayes2021deriving} for the complete details of single-reference expressions and the definition of $\Eval{e}{\sigma}$.}
\begin{align}
  \rely{r} \together \Expr{e}{k} = \rely{r} \together \Idle \Seq \cgd{(\EqEvalC{e}{k})} \Seq \Idle. \labelprop{single-reference}
\end{align}

\begin{lemmax}[reify-expr]
If $e_1$ is single reference under $r_1$, $e_2$ is single reference under $r_2$,
and $\prer{\CI} \inter r_1 \inter \postr{\CI} \subseteq r_2$ and
$\EqEvalC{e1}{k1} \supseteq \CI \inter \EqEvalC{e2}{k2}$ then,
\[
  \rely{r_1} \together \Expr{e1}{k1} \drefsto{\CI} \rely{r_2} \together \Expr{e2}{k2} .
\]
\end{lemmax}

\begin{proof}
The proof uses \refprop{single-reference}, then reifies the rely \refprop{reify-rely} and the test \refprop{reify-test}.
\begin{align*}
  \rely{r_1} \together \Expr{e1}{k1}
& = 
  \rely{r_1} \together \Idle \Seq \cgd{(\EqEvalC{e1}{k1})} \Seq \Idle \\
& \drefsto{\CI} 
  \rely{r_2} \together \Idle \Seq \cgd{(\EqEvalC{e2}{k2})} \Seq \Idle \\
& = 
  \rely{r_2} \together \Expr{e2}{k2}
 \qedhere
\end{align*}
\end{proof}

\begin{lemmax}[reify-conditional]
If $b_1$ and $b_2$ are boolean expressions that are single-reference under $r_1$ and $r_2$, respectively,
and $\prer{\CI} \inter r_1 \inter \postr{\CI} \subseteq r_2$
and $B_1 = \{ \sigma \spot (b_1)_{\sigma} \}$ and $B_2 = \{ \sigma \spot (b_2)_{\sigma} \}$,
and  $\CI \inter B_1 = \CI \inter B_2$, 
and both
$c_1 \drefsto{\CI} d_1$ and $c_2 \drefsto{\CI} d_2$,
\[
  \rely{r_1} \together \If b_1 \Then c_1 \Else c_2 \drefsto{\CI} \rely{r_2} \together \If b_2 \Then d_1 \Else d_2
\]
\end{lemmax}

\begin{proof}
The proof uses the definition of a conditional \refdef{conditional}, 
then  \reflem*{reify-expr} as $\CI \inter B_1 = \CI \inter B_2$,
and \refprop{reify-nondet} and \refprop{reify-seq}.
\begin{align*}
  \rely{r_1} \together \If b_1 \Then c_1 \Else c_2
& = 
  \rely{r_1} \together (\Expr{b_1}{\True} \Seq c_1 \nondet \Expr{b_1}{\False} \Seq c_2) \Seq \Idle \\
& \drefsto{\CI} 
  \rely{r_2} \together (\Expr{b_2}{\True} \Seq d_1 \nondet \Expr{b_2}{\False} \Seq d_2) \Seq \Idle \\
& = 
  \rely{r_2} \together \If b_2 \Then d_1 \Else d_2 
 \qedhere
\end{align*}
\end{proof}

\begin{lemmax}[reify-while]
If $b_1$ and $b_2$ are boolean expressions that are single-reference under $r_1$ and $r_2$, respectively,
and $\prer{\CI} \inter r_1 \inter \postr{\CI} \subseteq r_2$
and $B_1 = \{ \sigma \spot (b_1)_{\sigma} \}$ and $B_2 = \{ \sigma \spot (b_2)_{\sigma} \}$,
and $\CI \inter B_1 = \CI \inter B_2$, 
and 
$c_1 \drefsto{\CI} c_2$,
\[
  \rely{r_1} \together \While b_1 \Do c_1 \drefsto{\CI} \rely{r_2} \together \While b_2 \Do c_2
\]
\end{lemmax}

\begin{proof}
The proof is similar to that for \reflem{reify-conditional}.
\end{proof}

\section{Data refinement of a local variable block}\labelsect{reify-var}

The following lemma eliminates the abstract variable, $v$, except for ensuring $v$ is not changed by program transitions.
It is required to prove \Theorem{data-reification}, our main theorem.
\begin{lemmax}[diminish-inv]
If $v \neq w$ and $\id{w,v} \subseteq (\prer{I} \implies \postr{I})$ then 
\begin{displaymath}
\begin{array}{lll}
  & \Local{v}{\cgd{\CI} \Seq \inv{\CI}} \together \demand{\id{w}} \\
\refsto~& \cgd{(\HideS{v}{\CI})} \Seq \guar{(\prer{(\HideS{v}{\CI})} \implies \postr{(\HideS{v}{\CI})})}
    \together
    \demand{\id{w}}
    \together 
    \guar{\id{v}} ~.
\end{array}
\end{displaymath}
\end{lemmax}

\begin{proof}
\begin{align*}&
\Local{v}{\cgd{\CI} \Seq \inv{\CI}} \together \demand{\id{w}}
\Equals*[using invariant property \refprop{inv-maintains}, distributing $\demand{\id{w}}$ by \refprop{local-conj-indep} as independent of $v$]
\Local{v}{\cgd{\CI} \Seq \Om{\cstep{(\prer{I} \implies \postr{I})}} \together \demand{\id{w}}}
\Equals*[unfolding the definition of a local block \refdef{local}, and merging environment guarantees]
\guar{\id{v}} \together
\HideC{v}{(
  \cgd{\CI} \Seq \Om{\cstep{(\prer{I} \implies \postr{I})}}
  \together
  \demand{\id{w,v}}
)}
\Equals*[rewriting the atomic step iteration]
\guar{\id{v}} \together
\HideC{v}{(
  (\cgd{\CI} \Seq
   (\guar{(\prer{I} \implies \postr{I})}
   \together
   \demand{(\prer{I} \implies \postr{I})}
  )
  \together
  \demand{\id{w,v}}
)}
\Equals*[distributing test, and merging environment guarantees]
\guar{\id{v}} \together
\HideC{v}{(
  \cgd{\CI} \Seq
  ( \guar{(\prer{I} \implies \postr{I})}
    \together
    \demand{((\prer{I} \implies \postr{I}) \inter \id{w,v})}
  )
)}
\Refsto*[refining guarantee and applying assumption $\id{v,w} \subseteq (\prer{I} \implies \postr{I})$]
\guar{\id{v}} \together
\HideC{v}{(
  \cgd{\CI} \Seq
  ( \guar{(\prer{(\HideS{v}{\CI})} \implies \postr{I})}
    \together
    \demand{\id{w,v}}
  )
)}
\Equals*[by \refprop{hide-seq-iter-atomic} as
  $\HideLastC{v}{(\cpstep{(\prer{(\HideS{v}{\CI})} \implies \postr{I})} \nondet \cestep{\id{w,v}})}
  =   \cpstep{(\prer{(\HideS{v}{\CI})} \implies \postr{(\HideS{v}{\CI})})} \nondet \cestep{\id{w}}$
]
\guar{\id{v}} \together
\HideC{v}{(\cgd{\CI})}
\Seq
( \guar{(\prer{(\HideS{v}{\CI})} \implies \postr{(\HideS{v}{\CI})})}
  \together
  \demand{\id{w}}
)
\Equals*[distributing test over non-aborting commands, and commuting $\together$]
\HideC{v}{(\cgd{\CI})} \Seq \guar{(\prer{(\HideS{v}{\CI})} \implies \postr{(\HideS{v}{\CI})})}
\together \demand{\id{w}}
\together \guar{\id{v}}
\end{align*}
Where, for the application of property \refprop{hide-seq-iter-atomic},
we use the fact that a weak conjunction of program and environment guarantees can be rewritten as the iteration of an atomic step:
$\guar{g} \together \demand{r} = \Om{(\cpstep{g} \nondet \cestep{r})}$ 
for any $g$ and $r$.
\end{proof}

Our main theorem shows that a program using the abstract state is refined by the same program using the implementation state
via the coupling invariant $\CI$.
\begin{theoremx}[data-reification]
Given 
disjoint sets of variables $u$, $v$ and $w$, and 
coupling invariant $\CI$ relating $v$ and $w$, 
sets of (initial) states $iv$ and $iw$, 
commands $cv$, $cw$ and $dw$, 
if $iv$ and $cv$ are independent of $w$,
and
$iw$, $cw$ and $dw$ are independent of $v$,
and if
\begin{align}
  iw \inter \CI & \subseteq iv \labelprop{dr-init} \\
  \Pre{iw} \Seq \Frame{v,w,u}{cv} \together \demand{\id{v,w}} & \drefsto{\CI} \Frame{v,w,u}{cw} \together \demand{\id{v,w}} \labelprop{dr-CI} \\
  \id{w,v} & \subseteq (\prer{\CI} \implies \postr{\CI})  \labelprop{dr-cindep} \\
  %
  iw & \subseteq \HideS{v}{\CI} \labelprop{dr-iw-CI} \\
  \guar{(\prer{(\HideS{v}{\CI})} \implies \postr{(\HideS{v}{\CI})})} \together \Pre{iw} \Seq \Frame{w,u}{cw} & \refsto \Frame{w,u}{dw} \labelprop{dr-cw-CI}
\end{align}
hold then
\begin{math}
  \variable{v}{\Spec{v}{}{\postr{iv}} \Seq \Frame{u,v}{cv}} \refsto \variable{w}{\Spec{w}{}{\postr{iw}} \Seq \Frame{u,w}{dw}}~.
\end{math}
\end{theoremx}
\noindent
Given a coupling invariant $\CI$, $\HideS{v}{\CI}$ is the data-type invariant on the implementation variables $w$.
Assumption \refprop{dr-init} ensures the concrete initialisation correspond to an abstract initialisation;
\refprop{dr-CI} corresponds to reifying the body of the block;
\refprop{dr-cindep} ensures the coupling invariant $\CI$ is maintained if neither $v$ nor $w$ is modified;
\refprop{dr-iw-CI} ensures the initial state satisfies the data type invariant;
and 
\refprop{dr-cw-CI} eliminates the data type invariant.

\begin{proof}
\begin{align*}&
  \variable{v}{\Spec{v}{}{\postr{iv}} \Seq \Frame{v,u}{cv}}
 \Equals*[introduce local variable $w$ by \refprop{intro-local-framed} $w \neq v$; $iv$ and $cv$ are independent of $w$]
  \variable{v}{\Local{w}{\Spec{v,w}{}{\postr{iv}} \Seq \Frame{v,w,u}{cv}}}
 \Equals*[interchange variables $v$ and $w$ by \refprop{interchange-vars}]
  \variable{w}{\Local{v}{\Spec{v,w}{}{\postr{iv}} \Seq \Frame{v,w,u}{cv}}}
 \Refsto*[refine initialization using \refprop{dr-init} and $\Spec{v,w}{}{\postr{iw}} \Seq \cgd{\CI} = \Spec{v,w}{}{\postr{iw}} \Seq \cgd{\CI} \Seq \Pre{iw}$]
  \variable{w}{\Local{v}{\Spec{v,w}{}{\postr{iw}} \Seq \cgd{\CI} \Seq \cgd{iw} \Seq \Frame{v,w,u}{cv}}}
 \Equals*[restrict the scope of $v$ by \refprop{restrict-local} as $v \neq w$ and $iw$ is independent of $v$]
  \variable{w}{\Spec{w}{}{\postr{iw}} \Seq \Local{v}{\cgd{\CI} \Seq \cgd{iw} \Seq \Frame{v,w,u}{cv}}}
 \Refsto*[introduce coupling invariant $\CI$ by \refprop{inv-introduce} as $\cgd{\CI} = \cgd{\CI} \Seq \Pre{\CI}$]
  \variable{w}{\Spec{w}{}{\postr{iw}} \Seq \Local{v}{\cgd{\CI} \Seq (\inv{\CI} \together \cgd{iw} \Seq \Frame{v,w,u}{cv})}}
 \Refsto*[data refine using assumption \refprop{dr-CI}]
  \variable{w}{\Spec{w}{}{\postr{iw}} \Seq \Local{v}{\cgd{\CI} \Seq (\inv{\CI} \together \cgd{iw} \Seq \Frame{v,w,u}{cw})}}
 \Equals*[by \refprop{test-command-sync-command} assuming $\cgd{iw}\Seq \Frame{v,w,u}{cw}$ is not immediately aborting]
  \variable{w}{\Spec{w}{}{\postr{iw}} \Seq \Local{v}{\cgd{\CI} \Seq \inv{\CI} \together \cgd{iw} \Seq \Frame{v,w,u}{cw}}}
 \Equals*[by \refprop{local-conj-indep-framed} as $\Frame{u,v,w}{cw}$ is independent of $v$ and $\Spec{w}{}{\postr{iw}}$ establishes $\cgd{iw}$]
  \variable{w}{\Spec{w}{}{\postr{iw}} \Seq (\Local{v}{\cgd{\CI} \Seq \inv{\CI}} \together \Frame{w,u}{cw})}
 \Equals*[implicit $\demand{\id{w}}$ from local block for $w$ into local block for $v$]
  \variable{w}{\Spec{w}{}{\postr{iw}} \Seq (\Local{v}{\cgd{\CI} \Seq \inv{\CI}} \together \demand{\id{w}} \together \Frame{w,u}{cw})}
  \Equals*[by \reflem{diminish-inv} using \refprop{dr-cindep} and $v \neq w$]
  \variable{w}{
    \Spec{w}{}{\postr{iw}} \Seq
    (\cgd{(\HideS{v}{\CI})} \Seq \guar{(\prer{(\HideS{v}{\CI})} \implies \postr{(\HideS{v}{\CI})})}
    \together \demand{\id{w}}
    \together \guar{\id{v}} 
    \together \Frame{w,u}{cw} )
  }
 \Equals*[as $\guar{\id{v}} \together \Frame{w,u}{cw} = \Frame{w,u}{cw}$ and \refprop{dr-iw-CI}]
 \variable{w}{\Spec{w}{}{\postr{iw}} \Seq
    (\guar{(\prer{(\HideS{v}{\CI})} \implies \postr{(\HideS{v}{\CI})})}
    \together \demand{\id{w}}
    \together \Frame{w,u}{cw})}
 \Equals*[as $\demand{\id{w}}$ is implicit for local variable $w$ and \refprop{dr-cw-CI}]
  \variable{w}{\Spec{w}{}{\postr{iw}} \Seq \Frame{w,u}{dw}}
 \qedhere
\end{align*}
\end{proof}

\section{Example}\labelsect{example}

We consider a concurrent implementation of the Galler-Fischer data structure for implementing an equivalence relation \cite{1964GallerFischer}.
Our theory is developed in terms of sets of states and relations between states
but when presenting an example it is preferable to use predicates characterising both the sets of states and relations.
We use the notation $\Rel{pred}$ to denote the relation characterised by the predicate $pred$.

\paragraph{Specification.}

The abstract state for the specification is an equivalence relation $eq \subseteq X \times X$, 
for some finite set $X$, with $eq$ initially the identity relation.
All operations rely that their environment only grows the equivalence relation (via $equates$),
i.e.\ $eq_0 \subseteq eq$, where a $0$-subscripted variable refers to its initial value and and unsubscripted variable to its final value.
The specification of the operation $test$ is non-deterministic 
to allow for the fact that concurrent equates may take place while the test is executing:
$test(x,y)$ \underline{must} return $true$ if $x$ and $y$ are equivalent in the initial relation $eq_0$,
and \underline{may} return $true$ if $x$ and $y$ are relation in the final equivalence relation $eq$.
The operation $equate(x,y)$ ensures that $x$ and $y$ are in the equivalence relation after the operation.
In order to allow for concurrent test and equates, the equating of $x$ and $y$ must occur atomically,
and hence an atomic specification command is used.
The reflexive, transitive closure of a relation $r$ is written $\Fin{r}$.
Transitive closure is used in the atomic specification to ensure that for any $u$ and $z$, 
if $(u,x)$ and $(y,z)$ are in $eq_0$ then $(u,z)$ is in $eq$.
\begin{displaymath}
\begin{array}[t]{l}
  t \leftarrow test(x : X, y : X) \defs \\
  ~~\Rrely{eq_0 \subseteq eq} \together {} \\
  ~~\RSpec{t}{}{((x,y) \in eq_0 \implies t) \land (t \implies (x,y) \in eq)} \\[1ex]
  equate(x : X, y : X) \defs \\
  ~~\Rrely{eq_0 \subseteq eq} \together {} \\
  ~~\Frame{eq}{\Ratomicrel{eq = (eq_0 \union \{(x,y),(y,x)\})^{\star}}} 
\end{array}
\begin{array}[t]{l}
  clean\_up(x : X) \defs \\
  ~~\Rrely{eq_0 \subseteq eq} \together {} \\
  ~~\Frame{\emptyset}{\Term}
  \end{array}
\end{displaymath}
Collette and Jones \cite{ColletteJones00a} introduced the concept of an \emph{evolution invariant} $r$,
which corresponds to $\rely{r} \together \guar{r}$.
For our example, $eq_0 \supseteq eq$ is an evolution invariant.
Our concurrent rely-guarantee algebra supports evolution invariants \cite{2023MeinickeHayesDistributive-TR}.

\paragraph{Implementation.}

The Galler-Fischer implementation data structure is a forest of trees represented as a parent array $f \in X \fun X$, 
with data type invariant that the only cycles are of the form $f[n] = n$.
Initially $f$ is the identity function.
We define the relation that $n$ is a descendant of $m$ by,
\begin{equation}
  n \lef{f} m \defs (n,m) \in \Fin{f} . \labeldef{lef}
\end{equation}
The data type invariant on $f$ corresponds to requiring the relation $(\lef{f})$ to be a partial order, that is, it is
(a) reflexive, 
(b) transitive, and
(c) anti-symmetric.
Both (a) and (b) follow directly from the definition (because it uses reflexive, transitive closure).
Property (c) is an additional constraint,
\[
  n \lef{f} m \land m \lef{f} n \implies n = m
\]
which avoids any non-trivial cycles in the forest,
i.e.\ the only cycles are of the form $f\,n = n$ for the root elements.
We make use of the following auxiliary definitions on the forest representation:
$n \eqf{f} m$ states that $n$ and $m$ are equivalent within the forest $f$, 
that is, they have a common ancestor and hence are both within the same tree;
$roots\,f$ gives the set of all elements in $X$ that are the root of a tree;
and
$root\,f\,n$ gives the unique root of the tree within $f$ that $n$ is an element of.
\begin{align}
  n \eqf{f} m & \defs \exists k \spot n \lef{f} k \land m \lef{f} k \labeldef{eqf} \\
  roots\,f & \defs \{ n \in X \spot n = f\,n \} \labeldef{roots} \\
  root\,f\,n & \defs \If n = f\,n \Then n \Else root\,f\,(f\,n) \labeldef{root}
\end{align}
Note that $n \lef{f} m$ is a partial order, 
so $n \eqf{f} m$ is not the same as $n \lef{f} m \land m \lef{f} n$, 
as the later implies $n = m$ while the former only says $n$ and $m$ are both in the same tree.
From these definitions we can deduce the following properties.
\begin{align}
  (root\,f\,n = root\,f\,m) & \iff (n \eqf{f} m) \labelprop{roots-to-eqf} \\
  (n \lef{f} root\,f\,m) & \implies (n \eqf{f} m) \labelprop{root-eqf}
\end{align}
The function, $retr$, retrieves the equivalence relation from the forest.
\begin{equation}
  retr\,f \defs \{ (n,m) \in X \times X \spot n \eqf{f} m \} \labeldef{retr} 
\end{equation}
Our coupling invariant combines the retrieve function with the data type invariant.
\begin{equation}
  \CI \defs \Rel{eq = retr\,f \land partial\_order(\lef{f}) \labeldef{CI}}
\end{equation}

For the variant functions, $distance\,f\,x\,y$ is the number of arcs on the path from $x$ to $y$ within $f$,
assuming $x \lef{f} y$. 
The variant functions are in the form of a lexicographically ordered pair,
with the first component $\#(roots\,f)$ handling the case when there is interference that decreases the number of roots (via equates),
and the second component then handles the case where the number of roots does not change. 
\begin{align*}
 \begin{array}{l}
  rx : X \leftarrow  root\_of(x : X) \\
  ~~~~\begin{array}[b]{l}
            rx := x \Seq \\
            \While rx \neq f[rx] \Do \Variant (\#(roots\,f), distance\,f\,rx\,(root\,f\,rx)) \\
            ~~~~rx := f[rx] 
        \end{array} 
\\[1.5ex]
  t : \bool \leftarrow  test(x : X, y : X) \Var rx, ry : X;  \\
  ~~~~\begin{array}[b]{l}
           (rx := x \parallel ry := y) \Seq \\
           \While rx \neq f[rx] \lor ry \neq f[ry] \Do {} \\
           ~~~~\begin{array}{l}
                    \Variant (\#(roots\,f), distance\,f\,rx\,(root\,f\,rx)+distance\,f\,ry\,(root\,f\,ry)) \\
                    (rx := root\_of(rx) \parallel ry := root\_of(ry)) \Seq
                  \end{array} \\
           t := rx = ry
          \end{array} 
\\[1.5ex]
  equate(x : X, y : X) \Var rx,ry : X; done : \bool;  \\
  ~~~~\begin{array}{l}
           (rx := x \parallel ry := y \parallel done := false) \Seq \\
           \While \lnot done \Do \Variant (\#(roots\,f), distance\,f\,rx\,(root\,f\,rx)) \\
           ~~~~\begin{array}{l}
                    (rx := root\_of(rx) \parallel ry := root\_of(ry)) \Seq \\
                    done := CAS(f[rx],rx,ry) ~~~\Comment{or the symmetric version} \\
                   \end{array} 
          \end{array} 
 \end{array}
\end{align*}
\noindent
The implementation also introduces an operation $clean\_up(x)$ 
that reduces the length of the path from $x$ to its corresponding root, without changing the equivalence relation.
\begin{equation}
  clean\_up(x : X) \defs \Rrely{eq_0 \subseteq eq} \together \Frame{\emptyset}{\Term} \labeldef{clean-up}
\end{equation}
While multiple $test$ and $equate$ operations may run concurrently,
we assume that there is only one $clean\_up$ running at any one time. 

\paragraph{Reifying $clean\_up$.}

The first step in reifying the $clean\_up$ operation \refdef{clean-up} is to add the implementation state, $f$, to its frame, so it becomes,
\begin{equation}
  \Rrely{eq_0 \subseteq eq} \together \Frame{f}{\Term} . 
\end{equation}
We then reify this with coupling invariant \refdef{CI} above using \refprop{reify-rely} and \refprop{reify-frame} to give,
\begin{equation}
  \Rrely{retr\,f_0 \subseteq retr\,f} \together \Rguar{retr\,f_0 = retr\,f} \together \Frame{eq,f}{\Term} .
\end{equation}
The proof obligations are straightforward.
The rely condition allows,
for example, modifying $f_0$ for which $f_0\,n = m \land f_0\,m = m$ to $f$ with $f\,n = n \land f\,m = n$
but such a modification invalidates the assertion $n \not\in roots\,f$ (as well as others).
For the implementation of $clean\_up$ one needs a stronger rely condition
(which implies $retr\,f_0 \subseteq retr\,f$),
\begin{equation}
  \Rrely{roots\,f_0 \supseteq roots\,f \land (\forall n\,m \spot n \lef{f_0} m \implies n \lef{f} m)} . \labelprop{cleanup-rely}
\end{equation}
A standard rule in rely-guarantee theory is that a rely condition can be weakened
but here we need to strengthen the rely condition.
One can only strengthen this rely condition if all concurrent operations on $f$ guarantee this condition.
Here, this stronger rely condition \refprop{cleanup-rely} is guaranteed both by our implementation of $test$ 
(because it does not modify $f$)
and $equate$ (because it only ever adds to $f$).
Note, however, our implementation of $clean\_up$ does not guarantee \refprop{cleanup-rely},
which is why only a single $clean\_up$ at any one time is permitted.

The following gives an annotated version of the $clean\_up$ procedure.
The assertion just before the $\While$ loop is its invariant.
Variables $x$, $fx$ and $rx$ are local to $clean\_up$ and hence not subject to modification by other threads.
The auxiliary variable $f_0$ stands for the initial value of $f$.
\begin{displaymath}
\begin{array}{l}
  clean\_up(x : X) \Var fx,rx : X; \\
  ~~~\begin{array}{l}
          rx := root\_of(x) \Seq \\
          \SPre{root\,f_0\,x \lef{f_0} rx \lef{f} root\,f\,x \land x \lef{f} rx} \\
          \While x \neq rx \Do \Variant (\#(roots\,f), distance\,f\,x\,rx) \\
          ~~~~\begin{array}{l}
                   \SPre{root\,f_0\,x \lef{f_0} rx \lef{f} root\,f\,x \land x \ltf{f} rx} \\
                   fx := f[x] \Seq \\
                   \SPre{root\,f_0\,x \lef{f_0} rx \lef{f} root\,f\,x \land x \ltf{f} fx \lef{f} rx} \\
                   f[x] := rx \Seq \\
                   \SPre{root\,f_0\,x \lef{f_0} rx \lef{f} root\,f\,x \land fx \lef{f} rx} \\
                   x := fx \\
                   \SPre{root\,f_0\,x \lef{f_0} rx \lef{f} root\,f\,x \land x \lef{f} rx}
                   \end{array}
        \end{array}
 \end{array}
\end{displaymath}
One can show the assertions in the above program are stable under the strengthened rely.
The guarantee of $clean\_up$, that is, $retr\,f_0 = retr\,f$, 
is trivially satisfied by all commands other than the assignment to $f[x]$ because they do not modify $f$.
The assignment to $f[x]$ must satisfy the guarantee, that is, $retr\, f = retr\,(f[x \mapsto rx])$,
where $f[x \mapsto rx]$ is the same as $f$ except it maps $x$ to $rx$ (instead of $f\,x$).
Expanding the definition of $retr$ \refdef{retr}, this corresponds to showing using \refprop{roots-to-eqf},
\begin{align*}&
  \forall n\,m \spot (n \eqf{f} m) \iff (n \eqf{f[x \mapsto rx]} m)
 \Equiv
   \forall n\,m \spot (root\,f\,n = root\,f\,m) \iff  (root\,(f[x \mapsto rx])\,n = root\,(f[x \mapsto rx])\,m)
\end{align*}
If the paths $n$ and $m$ to their roots do not contain $x$, the equivalence holds trivially. 
If the path from $n$ to its root contains $x$, 
we have $root\,f\,n = root\,f\,x$ and $root\,(f[x \mapsto rx])\,n = root\,f\,rx = root\,f\,x$
from the precondition $rx \lef{f} root\,f\,x$ using \refprop{root-eqf},
and similarly if the path from $m$ to its root contains $x$,
and hence the equivalence is maintained.

\section{Conclusion}

Data reification is a powerful technique for separating the concerns of 
specifying a data structure (or type) in terms of an abstract representation
from 
the details of the concrete representation used to implement that data structure efficiently.
The rely-guarantee approach makes the task of reasoning about concurrent programs more tractable 
by providing a compositional approach that allows one to reason about a component of a program in relative isolation.

The approach we have taken is to extend a concurrent rely-guarantee algebra to handle data reification.
Our earlier research on generalised invariants \cite{2023MeinickeHayesDistributive-TR}, 
and in particular their distribution properties,
provides the basis for data reifying commands with respect to a coupling invariant.
Our research on localisation \cite{FormaliSE_localisation} provides a foundation for the proof of \Theorem{data-reification}
that justifies that a reification of a (hidden) data structure 
guarantees that any program using the implementation data structure 
is a refinement of the same program using the abstract specification of the data structure.

The approach of Liang, Feng and Fu \cite{LiangFengFuSimulation14} is also based on the rely-guarantee approach.
It uses a simulation technique, RGSim, 
that focuses on comparing externally observable behaviours only.

\bibliographystyle{splncs04}
\bibliography{ms}

\begin{thebibliography}{10}
\providecommand{\url}[1]{\texttt{#1}}
\providecommand{\urlprefix}{URL }
\providecommand{\doi}[1]{https://doi.org/#1}

\bibitem{Aczel83}
Aczel, P.H.G.: On an inference rule for parallel composition (1983), private
  communication to Cliff Jones
  \url{http://homepages.cs.ncl.ac.uk/cliff.jones/publications/MSs/PHGA-traces.pdf}

\bibitem{ColletteJones00a}
Collette, P., Jones, C.B.: Enhancing the tractability of rely/guarantee
  specifications in the development of interfering operations. In: Plotkin, G.,
  Stirling, C., Tofte, M. (eds.) Proof, Language and Interaction, chap.~10, pp.
  277--307. MIT Press (2000)

\bibitem{DaSMfaWSLwC}
Colvin, R.J., Hayes, I.J., Meinicke, L.A.: Designing a semantic model for a
  wide-spectrum language with concurrency. Formal Aspects of Computing
  \textbf{29},  853--875 (2016). \doi{10.1007/s00165-017-0416-4}

\bibitem{Dijkstra75}
Dijkstra, E.W.: Guarded commands, nondeterminacy, and a formal derivation of
  programs. CACM  \textbf{18},  453--458 (1975)

\bibitem{Dijkstra76}
Dijkstra, E.W.: A Discipline of Programming. Prentice-Hall (1976)

\bibitem{MPC19CylindricAlgebra}
Dongol, B., Hayes, I.J., Meinicke, L.A., Struth, G.: Cylindric {Kleene}
  lattices for program construction. In: Hutton, G. (ed.) Mathematics of
  Program Construction 2019. Lecture Notes in Computer Science, vol. 11825.
  Springer International Publishing, Cham (Oct 2019).
  \doi{10.1007/978-3-030-33636-3_8}

\bibitem{1964GallerFischer}
Galler, B.A., Fischer, M.J.: An improved equivalence algorithm. Commun. ACM
  \textbf{7}(5),  301–303 (May 1964). \doi{10.1145/364099.364331},
  \url{https://doi.org/10.1145/364099.364331}

\bibitem{HJM-23}
Hayes, I.J., Jones, C.B., Meinicke, L.A.: Specifying and reasoning about
  shared-variable concurrency. In: Bowen, J.P., Li, Q., Xu, Q. (eds.) Theories
  of Programming and Formal Methods. pp. 110--135. No. 14080 in LNCS, Springer
  (2023). \doi{10.1007/978-3-031-40436-8_5}

\bibitem{hayes2021deriving}
Hayes, I.J., Meinicke, L.A., Meiring, P.A.: Deriving laws for developing
  concurrent programs in a rely-guarantee style (2021).
  \doi{10.48550/ARXIV.2103.15292}, \url{https://arxiv.org/abs/2103.15292}

\bibitem{HenkinMonkTarski71}
Henkin, L., Monk, J.D., Tarski, A.: Cylindric Algebras, Part I, Studies in
  logic and the foundations of mathematics, vol.~64. North-Holland Pub. Co.
  (1971)

\bibitem{Hoare72-data}
Hoare, C.A.R.: Proof of correctness of data representations. Acta Informatica
  \textbf{1},  271--281 (1972), also in Programming Methodology, D. Gries (ed)
  Springer-Verlag (1978)

\bibitem{Jones80a}
Jones, C.B.: Software Development: A Rigorous Approach. Prentice Hall
  International (1980), \url{http://portal.acm.org/citation.cfm?id=539771}

\bibitem{Jones81d}
Jones, C.B.: Development Methods for Computer Programs including a Notion of
  Interference. Ph.D. thesis, Oxford University (June 1981), available as:
  Oxford University Computing Laboratory (now Computer Science) Technical
  Monograph PRG-25

\bibitem{Jones83a}
Jones, C.B.: Specification and design of (parallel) programs. In: Proceedings
  of IFIP'83. pp. 321--332. North-Holland (1983)

\bibitem{Jones83b}
Jones, C.B.: Tentative steps toward a development method for interfering
  programs. ACM ToPLaS  \textbf{5}(4),  596--619 (1983).
  \doi{10.1145/69575.69577}

\bibitem{Jones90a}
Jones, C.B.: Systematic Software Development using VDM. Prentice Hall
  International, second edn. (1990),
  \url{http://homepages.cs.ncl.ac.uk/cliff.jones/ftp-stuff/Jones1990.pdf}

\bibitem{Jones06a}
Jones, C.B.: Splitting atoms safely. Theoretical Computer Science
  \textbf{375}(1--3),  109--119 (2007)

\bibitem{LiangFengFuSimulation14}
Liang, H., Feng, X., Fu, M.: Rely-guarantee-based simulation for compositional
  verification of concurrent program transformations. ACM Transactions on
  Programming Languages and Systems  \textbf{36}(1),  3:1--3:55 (2014)

\bibitem{FormaliSE_localisation}
Meinicke, L.A., Hayes, I.J.: Using cylindric algebra to support local variables
  in rely/guarantee concurrency. In: 2023 IEEE/ACM 11th International
  Conference on Formal Methods in Software Engineering (FormaliSE). pp.
  108--119. IEEE (2023). \doi{10.1109/FormaliSE58978.2023.00019}

\bibitem{2023MeinickeHayesDistributive-TR}
Meinicke, L.A., Hayes, I.J.: Reasoning about distributive laws in a concurrent
  refinement algebra (2024),
  \href{https://arXiv.org/abs/arXiv:2403.13425}{arXiv:2403.13425} [cs.LO]

\bibitem{Milner71a}
Milner, R.: An algebraic definition of simulation between programs. Tech. Rep.
  CS-205, Computer Science Dept, Stanford University (February 1971)

\bibitem{IsabelleHOL}
Nipkow, T., Paulson, L.C., Wenzel, M.: Isabelle/HOL: A Proof Assistant for
  Higher-Order Logic, LNCS, vol.~2283. Springer (2002)

\end{thebibliography}

\end{document}